\documentstyle[preprint,eqsecnum,aps,epsfig]{revtex}

\begin{document}
\draft
\title{{QCD near the Light Cone}}
\author{H. W. L. Naus$^{a,b}$, H. J. Pirner$^a$, T. J. Fields$^{a,c}$, 
and J. P. Vary$^{a,c}$}
\address{
$^a$Institute for Theoretical Physics, University of Heidelberg \\
    Philosophenweg 19, 69120 Heidelberg, Germany \\
$^b$Institute for Theoretical Physics, University of Hannover \\
    Appelstr. 2, 30167 Hannover, Germany \\
$^c$Department of Physics and Astronomy, Iowa State University \\
    Ames, IA  50011}

\date{\today}
\maketitle

\begin{abstract}
Starting from the QCD Lagrangian, we present the QCD Hamiltonian
for  near light cone coordinates. We study the 
dynamics of the gluonic zero modes of this Hamiltonian.
The strong coupling solutions serve as a basis for
the complete problem. We discuss the importance of zero
modes for the confinement mechanism.
\end{abstract}

\section{Introduction}

The formulation of light front QCD is one of the most innovative 
enterprises in recent theoretical hadron physics. 
It resumes the pioneering
efforts of the seventies in the parton model \cite{BKS,KS}. 
Its intention
is to connect the successful parton model at large resolution 
$Q^2$ with the
constituent quark picture of hadrons appearing in  spectroscopy. The 
new start  \cite{PB} is
not without a knowledge of the problems which have been 
experienced in the first
works. It is well documented that renormalization in Hamiltonian 
field theories is, with currently
available methods,  
more cumbersome than in covariant descriptions. A naive gauge fixing 
procedure on the light front leads to an easy resolution of Gauss' Law.
However, this naive method is not correct -- the correct method involves
so--called `zero mode' degrees of freedom dependent on the transverse 
coordinates.  These zero modes cannot be gauged away, and become an
integral part of the dynamics.  In addition we expect that the 
nontrivial vacuum structure 
evident from equal time quantization brings  new induced couplings
into the light front Hamiltonian.

Our investigation concentrates on the role of the zero mode fields.
We start from the near light front 
frame advocated by
the St. Petersburg and Erlangen groups \cite{PRFR,LTLY} and 
introduce a  finite volume ($L \times L \times L$ in the spatial
directions), thereby controlling possible infrared singularities.
This choice of coordinate system involves quantization on a space--like
surface, which makes it easier to relate the occurring phenomena
to equal time physics. 
However, all the usual complexities of
negative energy states and nontrivial vacua are also present.

An axial gauge is natural in the infinite momentum frame: the axis
of motion singles out a preferred direction.
The chosen gauge $\partial_-A_-=0$ is actually a variant of the 
axial gauge, since the latter is incompatible with the boundary conditions.
Consequences include  the appearance and quantization of the zero--mode 
gauge fields $a_-(x_\perp)$ which depend only on the two transverse 
coordinates and can be chosen color diagonal.
The asymmetrical dependence of the zero modes on only the
transverse coordinates coincides with the 
asymmetry of the space coordinates in this near light front frame. 
Explicitly, the light front spatial $x^-$ direction stems
from mixing the former spatial $x^3$ coordinate  with the
former time $x^0$.

The zero modes degrees of freedom appear in a transverse Hamiltonian which is 
coupled to
three--dimensional dynamics via the fermions, transverse gauge 
fields and the
light front Coulomb law. In the strong coupling limit the kinetic term of 
the transverse Hamiltonian becomes dominant and,  
in this limit, the Hamiltonian is identical to a Hamiltonian 
describing independent rotators at each lattice site. 
This is the  starting point of our investigation.
Using basis functions according to 
this  dominant kinetic term, we make an expansion of the zero mode 
Hamiltonian.
In the strong coupling basis we also
evaluate the Coulomb term and find indications that the interaction of 
external sources is confining.

In Hamiltonian formulations quantized on the exact light front,
the zero modes $a_-(x_\perp)$  correspond to high energies.
They are infinitely long wavelength excitations with respect 
to the spatial variable $x^-$, and, simultaneously,  they are very short
wavelength (ultraviolet) excitations with respect to the time variable $x^+$.
Therefore some approaches \cite {WW} include the zero mode physics in their
renormalization program, without explicitly solving their dynamics.
In our approach of near light cone coordinates, 
the zero modes are independent degrees of freedom and retained 
after the solution of Gauss' Law,
since they correspond to  gauge invariant
quantities (as evident in their role as eigenphases of the Polyakov loops).
Resolution of the Gauss law constraint indeed 
does not permit the elimination
of their conjugate momenta $p^-(x_\perp)$.

The existence of zero
modes indicates that the local color charge of all external sources will be 
singlet -- all hadronic bound states must be  color 
singlets. This is the first requirement of color confinement. The 
second requirement for color confinement
dictates that the interaction energy increases with increasing separation.
To make a reliable calculation in this case, one has to know,  besides  the
light front Coulomb potential, 
how gluon fields propagate 
in the `background' field $a_{-}(x_\perp)$ over large distances.
It is argued that on a strongly coupled lattice the background field
modifies the behavior of transverse plane gluon waves with
color charge. Due to the fluctuations of the background they are 
limited to propagate over short distances. Thus they cannot 
cancel the linear confinement potential induced by the gauge choice.
In this way, the axial gauge supports confinement from the beginning and 
therefore seems to be the best starting point for QCD.
\label{confine}
However, confinement is not so obvious for neutral 
transverse gluon fields,  whose two-dimensional longitudinal
zero mode part has been eliminated  
in the procedure of solving the Gauss law constraint.

Light front gauge theories have been set up using the so--called
near light front frame \cite{PRFR,LTLY,PNP,VFP}, which enables one 
to study the approach to the exact light front. 
The tilted coordinates are defined as
\begin{eqnarray}
x^{+} & = & \frac{1}{\sqrt{2}} \left\{ \left(1+ \frac{\eta^2}{2} \right)
x^{0} + \left( 1 - \frac{\eta^2}{2} \right) x^{3} \right\}  \ ,\nonumber 
\\
x^{-} & = & \frac{1}{\sqrt{2}} \left( x^{0}-x^{3} \right) \ .
\label{Coor}
\end{eqnarray}
The transverse components $x^1$ and $x^2$ are unchanged;
$x^{+}$ is the new time coordinate and $x^{-}$ is the remaining
spatial coordinate. 
As finite quantization volume we will take a torus and its
extension in $x^-$, as well as in $x^1$, $x^2$ direction is $L$.
The metric tensor reads
\begin{equation}
g_{\mu \nu} = \left( \begin{array}{rrrr} 0 & 0 & 0 & 1 \\
0 & -1 & 0 & 0 \\ 0 & 0 & -1 & 0 \\ 1 & 0 & 0 & -\eta^2
\end{array} \right) \ \ , \ \ \ \
g^{\mu \nu} = \left( \begin{array}{rrrr} \eta^2 & 0 & 0 & 1 \\
0 & -1 & 0 & 0 \\ 0 & 0 & -1 & 0 \\ 1 & 0 & 0 & 0
\end{array} \right)  \ ,
\label{metric}
\end{equation}
where $\mu\ , \nu = +,1,2,-$.
It defines the scalar product of two 4-vectors $x$ and $y$:
\begin{eqnarray}
x_{\mu}y^{\mu} & = & x^{-}y^{+} + x^{+}y^{-} - \eta^2 x^{-}y^{-}
-  \vec{x}_{\bot}  \vec{y}_{\bot} \nonumber \\
               & = & x_{-}y_{+} + x_{+}y_{-} + \eta^2 x_{+}y_{+}
-  \vec{x}_{\bot}  \vec{y}_{\bot} \ ,
\label{scalpr}
\end{eqnarray}
where
\begin{equation}
 \vec{x}_{\bot}  \vec{y}_{\bot} = x^{1}y^{1} + x^{2}y^{2} \  .
\label{tran}
\end{equation}
Obviously, the exact light-cone is approached as
the parameter $\eta^2$ goes to zero.
For non--zero $\eta$, the transition to the near light front coordinates 
from an equal time frame can be formally identified as a Lorentz boost  
combined with a linear transformation which avoids time dependent boundary 
conditions \cite{LTLY}, as required in the canonical formulation. 
The boost parameter $\beta$ for this Lorentz transformation 
(in the $x^3$ direction) is given by
\begin{equation}
\beta = \frac{1 - \eta^2/2}{1+\eta^2/2} \, ,
\label{beta}
\end{equation}
indicating that for $\eta \rightarrow 0$ the relative velocity of the
two frames $ v = \beta c \rightarrow c\, (\equiv 1)$.  This is connected to the
interpretation of the near light front frame in terms of the infinite 
momentum frame.  As discussed previously, the use of these near 
light front coordinates allows us to quantize the theory on a spacelike
surface at equal light front time.  This is important, as the
ends of our `box' are spacelike separated, and can thus exchange
no information.  Note the contrast to conventional
light front coordinates, in which the ends of the `box' are
separated by a light--like interval which, in turn, implies that the
surface of quantization contains points which can be causally connected.

The spectrum of massless partons $p_{\mu}p^{\mu}=0$ 
with transverse momentum
$p_{\bot}$ can be easily calculated from their dispersion
relation which reads in the near 
light front coordinates
\begin{equation}
2 p_-p_+  + \eta^2 (p_+)^2-p_{\bot}^2=0.
\end{equation}
In our coordinates, $p_+$ is the energy variable
(being conjugate to $x^+$) and $p_-$ is the longitudinal momentum variable.  
This dispersion relation has two $p_+$ solutions, an upper 
branch $p_{+,\rm up}$
and a lower branch $p_{+,\rm down}$
\begin{equation}
p_{+,\rm up/down} = -\frac{1}{\eta^2}
\left( p_- \mp \sqrt{(p_-)^2 + \eta^2 p_{\bot}^2}\right).
\label{disp1}
\end{equation}
In the limit of small $\eta$, Eq. (\ref{disp1})
goes over into the conventional light front energy
\begin{equation} 
p_{+,\rm up} \rightarrow p_{+,\rm lc}=p_{\bot}^2/2p_{-,\rm lc}.
\label{plc}
\end{equation}

\begin{figure}[hbt]
\epsfig{figure=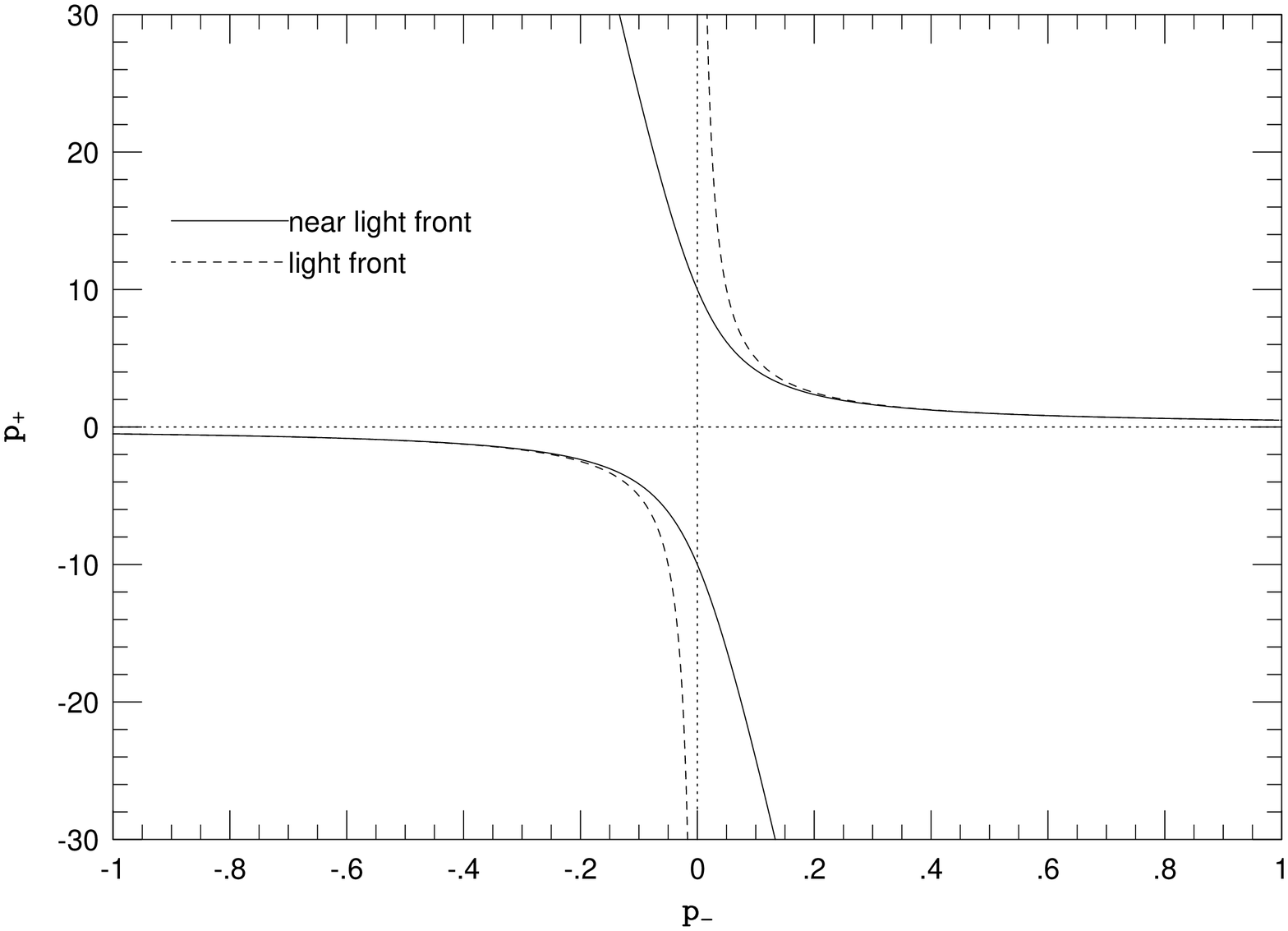, scale=0.5}
\caption{\label{disp} Dispersion relations in the near light front coordinate system
and directly on the light front.}
\end{figure}

In Figure \ref{disp}, we show these two branches together with the 
light front Hamiltonian $p_{+,\rm lc}=p_{\bot}^2/2p_{-,\rm lc}$.
We have taken the maximum transverse momentum $p_{\bot}=1/a$,
$\eta=0.1$, and all energies and momenta are in units of $1/a$,
where $a$ is the lattice spacing.  
Note that negative momentum states correspond uniquely
to negative energy states in the conventional light front formalism, 
but with our choice of coordinates, this clean division no longer holds.  
However, it
is true that for small $\eta$, the negative momentum states with
positive energy are of very high energy, indeed.   One possible 
procedure to find the momenta 
for regularization of these high energy states is to calculate the 
longitudinal momenta corresponding 
to a maximum  absolute energy
\begin{equation}
-\frac{1}{\eta a} < p_+ < \frac{1}{\eta a}.
\end{equation}
For $p_{\bot} = 1/a$, this cutoff corresponds
to $p_-=0$ in $p_{+,\rm up/down}$ and renders the
dispersion relation single--valued once again: a unique energy $p_+$
corresponds to each momentum $p_-$.

Since we are finally interested in using 
$p_{+,\rm lc}$ as our effective Hamiltonian,
the upper energy cutoff gives the minimal $p_-$ momentum
where this approximation still makes sense.
When we use the expression in Eq. (\ref{plc}) for  $p_{+,\rm lc}$ 
we obtain a minimal $p_-= \eta / 2a $, resulting from our energy
cutoff. 
To introduce an efficient light front effective theory, it is 
necessary to eliminate partons with negative  $p_-$ states
with special attention to the partons with $0< p_- < \eta / 2a $.
This procedure will not be discussed here, but we believe
that this elimination is very important in order to obtain a constituent quark 
picture on the light front.  We would further comment that our effective
Hamiltonian approach
discussed in the next chapter may provide a workable framework to
include these high energy modes.

This paper is divided in six sections. 
In the next section we review the QCD
Hamiltonian in the case of these tilted light front coordinates.
In section 3 we give the
solution to the zero mode sector in 
the strong and weak coupling approximations.
A calculation 
of the ground state energy in the two site approximation
is presented in section 4.
Section 5 contains a discussion of confinement.
Finally, we summarize in section 6.

\section{QCD Hamiltonian near the Light Front} 

Canonical formulations of QED in the axial gauge
and in the light front gauge have been developed in analogous ways --  
starting from the respective canonical Weyl gauges.
After the implementation of the Gauss law constraints,
the resulting Hamiltonians
appear to be rather similar \cite{LNOT}.
Moreover, the QCD Hamiltonian in axial gauge representation 
has recently been derived  \cite{LNT}. 
Here we will outline the derivation
of the near light front QCD Hamiltonian, which has been
given by the Erlangen group \cite{LNT2} also.
We restrict ourselves to the color gauge group $SU(2)$ and dynamical
gluons; only an
external (fermionic) charge density $\rho_m$ is considered here.
General $SU(N)$ results, including dynamical fermions, are
given in the Appendix.

The Lagrangian in the near light front coordinate system reads
\begin{equation}
{\cal L} = \frac{1}{2}F^a_{+-}F^a_{+-} + \sum_{i=1,2}
\left( F^a_{+i}F^a_{-i} + \frac{\eta^2}{2}F^a_{+i}F^a_{+i}\right)
-\frac{1}{2}F^a_{12}F^a_{12} - \rho_m^a A_+^a,
\end{equation}
where the color index $a$ is summed from 1 to 3, and the transverse
coordinates are labeled by $i=1,2$.
We will also use the  matrix notation; for example $ A_- = A_-^a \tau^a / 2 $,
where the $\tau^a$ are the $SU(2)$ matrices. The field strength tensor
contains the commutator of the gauge fields:
\begin{equation}
F_{\mu \nu} = \partial_{\mu} A_{\nu} -\partial_{\nu} A_{\mu}
-i g [A_{\mu}, A_{\nu}] \, ,
\end{equation}
with the coupling constant $g$.

The $A^a_+$ coordinates have no momenta conjugate to them. 
As a consequence, the
Weyl gauge $A^a_+=0$ is the most natural starting point for
a canonical formulation.
The canonical momenta of the dynamical fields $A_-^a, A_i^a$ are given by
\begin{eqnarray}
\Pi^a_- &=& \frac{\partial{\cal L}}{\partial F^a_{+-}} = F^a_{+-}, 
\nonumber \\
\Pi^a_i &=& \frac{\partial{\cal L}}{\partial F^a_{+i}} = F^a_{-i}
+ \eta^2 F^a_{+i}.
\end{eqnarray}
From this, we get the Weyl gauge Hamiltonian density
\begin{equation}
{\cal H}_W = \frac{1}{2} \Pi^a_- \Pi^a_- + \frac{1}{2}F^a_{12}F^a_{12}
+ \frac{1}{2\eta^2}\sum_{i=1,2}\left( \Pi^a_i - F^a_{-i}\right)^2.
\label{Hweyl}
\end{equation}
We choose periodic boundary conditions in $x^-$ and $x_{\bot}$
on intervals of size $[0,L]$.  Using the appropriate periodic
delta functions, the quantization is straightforward. However,
the Hamiltonian has to be supplemented by the original Euler--Lagrange 
equation for $A_+$ as  constraints on the physical states
\begin{eqnarray}
G^a({x}_{\bot},x^-) | \Phi \rangle &=&
\left( D_-^{ab} \Pi^b_{-} + D_{\bot}^{ab}\Pi^b_{\bot} + g \rho_m^a \right)
 | \Phi \rangle \nonumber \\
&=& \left( D_-^{ab} \Pi^b_{-} + G_{\bot}^{a}\right) | \Phi \rangle
=  0,
\label{Gauss}
\end{eqnarray}
with the covariant derivatives
\begin{eqnarray}
D_{-}^{ab} &=& \partial_{-}\delta^{ab} + g f^{acb} A_{-}^c , \nonumber \\ 
D_{{\bot}}^{ab} &=& \partial_{{\bot}}\delta^{ab} + g f^{acb} A_{{\bot}}^c ,
\label{Deriv}
\end{eqnarray}
where $f^{acb}$ are the structure constants of $SU(2)$.

These equations are known as Gauss' Law constraints. 
Since the Gauss' Law operator commutes with the Hamiltonian
\begin{equation}
[ G^a({x}_{\bot},x^-) ,  H_{W} ]    = 0 ,
\end{equation}
time evolution leaves the system in the space of physical states.
Furthermore, $H_{W}$ is invariant under time independent residual
gauge transformations whose  generator is 
closely connected to Gauss' Law \cite{LNT}.

In order to obtain a Hamiltonian formulated in terms of unconstrained
variables, thus 
rendered available for approximations without breaking local gauge
invariance,  one needs to resolve the Gauss' Law constraint. 
This is an  important step, since it frees us from implementing
additional restrictions on operators or states.  A Hamiltonian 
expressed in terms of unconstrained variables appears more complicated,
but it is our belief that there is much to gain from this 
apparent increase in complexity.  

Via unitary gauge fixing transformations \cite{LNOT,LNT} one indeed
can achieve this resolution with respect to components of
the chromo--electric field. These transformations render a Hamiltonian
independent of the conjugate gauge fields. In other words, the
latter become cyclic variables. 
Let us choose the `$-$' (minus) components as
the variables to be eliminated. Classically this would correspond to the
light front gauge $A_- =0$. However, this choice is not legitimate 
for our setup in a finite box.
Only the (classical)
Coulomb light front gauge ($\partial_{-}A_{-}=0$) is compatible
with gauge invariance and periodic boundary conditions.
The reason is that $A_-$ carries information on gauge invariant 
quantities, such as the eigenvalues of the spatial Polyakov (Wilson) loop 
\begin{equation}
{\cal P}({x}_\bot)=  P \exp\left[ ig\int dx^-A_-(x_{\bot},x^-) 
\right] \, ,
\end{equation}
which can be written in terms of a diagonal matrix
$a_-({x}_{\bot})$
\begin{equation}
{\cal P}({x}_\bot)= V \exp\left[ ig L a_-(x_{\bot}) \right] V^{\dagger} \, .
\end{equation}
Thus, we obviously need to keep these `zero modes' $a_-({x}_{\bot})$
as dynamical variables, while 
the other components of $A_-$ are eliminated.
The zero mode degrees of freedom are independent of
$x^-$ and, therefore, correspond to quantities with zero longitudinal momentum.

The unitary gauge fixing transformation can indeed be chosen
in such a way that $A_-$ becomes cyclic, apart from the zero modes
mentioned \cite{LNT2}. In order to eliminate the conjugate
momentum, $\Pi_-$, by means
of Gauss' Law,  one needs to `invert' the
covariant derivative $D_-$. After the unitary transformation $D_-$
simplifies significantly (compare to Eq. (\ref{Deriv}))
\begin{equation}
D_- \rightarrow d_- = \partial_- -ig \, [a_-,  \;\;\;\;\;\;\;  ,
\end{equation}
where this is understood as an operator equation: e.g. 
$D_-f \rightarrow d_-f = \partial_-f -ig \, [a_-,f]$ for arbitrary f.
Now Gauss' Law can be readily resolved: in the space of physical states
one can make the replacement 
\begin{equation}
\Pi_{-}({x}_{\bot},x_{-}) \rightarrow p_{-}({x}_{\bot}) - \left (
d_-^{-1} \right ) G_{\bot}({x}_{\bot},y^-) .
\end{equation}
The inversion of $d_-$ can be explicitly constructed in terms of its
eigenfunctions, cf. \cite{LNT,LNT2}.
The zero mode  operator $p_-({x}_{\bot})$ is also diagonal and 
$p^3_-({x}_{\bot})$ is the conjugate momentum
to $a^3_{- }(x_{\bot})$. It has eigenvalue zero with respect to $d_-$,
i.e. $d_- p_- =0$,
and is therefore not constrained.

The appearance of the zero modes
also implies residual Gauss' Law constraints. In the space of
transformed physical states $|\chi \rangle$, they read
\begin{equation}
\int dx^- G_{\bot}^3 \, |\chi \rangle =
\int dx^- \left ( D_{\bot}^{3b} \Pi_{\bot}^b + g\rho_m^3 \right ) 
|\chi \rangle =0.
\label{RG}
\end{equation}
These two--dimensional constraints can be handled in full analogy to 
QED, since they correspond to the diagonal part of color space. 
This further gauge fixing in the $SU(2)$ 3--direction is done via 
another gauge fixing transformation, which leads
to the Coulomb gauge representation in the transverse plane for the neutral
fields.  In other words, we eliminate the color neutral,
$x^-$--independent, two--dimensional longitudinal gauge fields
\begin{equation}
{a}_{\bot}^{\ell}(x_{\bot}) = \frac{1}{L} \int dy^- dy_{\bot}
d(x_{\bot}-y_{\bot}) \nabla_{\bot}
\left( \nabla_{\bot} \cdot A_{\bot}^3({y}_{\bot}, y_-) \right)
 \frac{\tau^3}{2} \,.
\label{SUB}
\end{equation}
Here we use the periodic Greens function
of the two dimensional Laplace operator
\begin{equation}
d({z}_{\bot}) = - \frac{1}{L^2} \sum_{\vec{n} \neq \vec{0} }
\frac{1}{p_n^2} e^{i p_n z_{\bot}}  \ ,
\ \ \ \ p_n = \frac{2 \pi}{L} \vec{n} \ ,
\label{Green2}
\end{equation}
where $\vec{n} = (n_1,n_2)$ and $n_1,n_2$ are integers.  
The conjugate momenta of these fields, 
$p_{\bot}^{\ell} (x_{\bot})$,
are defined analogously.
Resolution of the residual Gauss' Law allows one to replace them,
in the sector of the transformed physical space $| \Phi' \rangle$,
by the  neutral chromo--electric field
\begin{equation}
e_{\bot} (x_{\bot}) = g \nabla_{\bot}\int dy^{-}dy_{\bot}
d(x_{\bot} - y_{\bot})
\left \{ f^{3ab} A^a_{\bot}(y_{\bot}, y^-) \Pi^b_{\bot}(y_{\bot}, y^-)  
+ \rho^3_m(y_{\bot}, y^-)  \right \} \frac{\tau^3}{2} \, .
\end{equation}
At this point it is convenient to introduce
the unconstrained gauge fields and their conjugate momenta:
\begin{eqnarray}
A_{\bot}'(x_{\bot}, x^-) &=& 
A_{\bot}(x_{\bot},x^{-}) - a_{\bot}^{\ell}(x_{\bot}),
 \nonumber \\
\Pi_{\bot}'(x_{\bot}, x^-) &=& \Pi_{\bot}(x_{\bot}, x^-) - 
\frac{1}{L} p_{\bot}^{\ell}(x_{\bot}).
\label{subt}
\end{eqnarray}
These  relations turn out to be  important for  
neutral gluon exchange; recall that the subtracted fields
are diagonal in color space.
Note that the physical degrees of freedom $A_{\bot}'$ and
$\Pi_{\bot}'$ still contain $(x_{\bot}, x^-)$--independent,
color neutral, modes. Therefore, there is a 
remnant of the local Gauss' Law constraints -- the global
condition  
\begin{equation}
Q^3 |\Phi ' \rangle = 
\int dy^{-}dy_{\bot}
\left \{ f^{3ab} A^a_{\bot}(y_{\bot}, y^-) \Pi^b_{\bot}(y_{\bot}, y^-)  
+ \rho^3_m(y_{\bot}, y^-)  \right \} 
 |\Phi ' \rangle =0 \,. 
\label{Gaussgl}
\end{equation}
The physical meaning of this equation is 
that the neutral component of the total color 
charge, including external matter as well as gluonic contributions, must
vanish in the sector of physical states.  

The final Hamiltonian density in the physical sector explicitly reads
\begin{eqnarray}
{\cal H} & = & \mbox{tr} \left[  \partial_1 A'_2 -\partial_2 A'_1
-ig [A'_1, A'_2]  \right]^2 
 +  \frac{1}{\eta^2} \mbox{tr} \left[ 
\Pi'_{\bot}-\left(\partial_-A'_{\bot}-
ig[a_-,A'_{\bot}]\right)\right]^2 \nonumber \\
& + &  \frac{1}{\eta^2} \mbox{tr} \left[ \frac{1}{L} e_{\bot} 
-
\nabla_{\bot} a_-\right]^2 
 +  \frac{1}{2 L^{2}}  p_{-}^ 
{3 \, \dagger}(x_{\bot})
p^3_{-}(x_{\bot})  \nonumber \\
& + &   \frac{1}{L^{2}}\int_{0}^{L} dz^{-} \int_{0}^{L} dy^{-}
\sum_{p,q,n}\,^{'}\frac{ G'_{\bot qp}(x_{\bot}, z^{-})
G'_{\bot pq}(x_{\bot},y^{-})}{\left[ \frac{2\pi n}{L} +
g(a_{-q}(x_{\bot})- a_{-p}(x_{\bot})) \right]^{2}}
e^{i2\pi n(z^{-}-y^{-})/L} \ ,
\label{Hamil}
\end{eqnarray}
where $p$ and $q$ are matrix labels for rows and columns, 
 $a_{-q}=(a_-)_{qq}$ and the prime indicates that the summation is  
restricted to $n \neq 0$ if $p=q$.
The operator $G'_{\perp}\left(x_{\bot}, x^-\right)$ is defined as
\begin{equation}
G'_{\perp}=
\nabla_{\perp}\Pi'_{\perp}
+ gf^{abc}\frac{\tau^{a}}{2}A'\,^{b}_{\perp}
\left({\Pi'}\,^{c}_{\perp} -
\frac{1}{L}{e}\,^{c}_{\bot}\right) 
+ g\rho_{m}   \,.
\label{Gop}
\end{equation}

In Table 1 we show the number of degrees of freedom for the
different stages of the gauge fixing procedure. In order to
count, we discretize configuration space as $N^3$ sites.
Constraints of course reduce the number of independent variables
and are therefore subtracted.
Originally there are three vector components  
of the gluon field in three colors at each site. 
Gauss' law represents three (color) constraints at each site.
After the first unitary transformation we arrive at a intermediate
Hamiltonian which is not explicitly given in the text.
The number of degrees of freedom as well as the number of
constraints have been reduced. Note that the residual Gauss law
is two--dimensional and color neutral.
Finally, we arrive at a formulation where there is only
one global constraint left. Concomitantly, the final $A'_{\bot}$ still
contains the global color neutral zero mode which has not
been subtracted (cf. Eqs. (\ref{SUB}, \ref{Green2})).
Thus, one explicitly sees that the number of independent
unconstrained degrees of freedom is $6N^3$ at {\it any} stage
of the formal development.

\begin{table}
\begin{center}
\caption{Counting the degrees of freedom in the 
$SU(2)$ Hamiltonians}
\vspace{0.3in}
\begin{tabular}{|lcr|}
\hline
{\bf Original ${\cal H}$} &  &  \\
$A^a_1,A^a_2,A^a_-(x_\bot,x_-):$ &\hspace{0.5in} &  $3*3*N^3$ \\
Gauss' law constraints:   &\hspace{0.5in}
	&  $-3*N^3$ 	\\
 &\hspace{0.5in} & ------------------\\
&\hspace{0.5in} &  $6N^3$ \\
\hline
{\bf Intermediate ${\cal H}$} &  &  \\
$A_1^a, A_2^a(x_\bot,x_-):$  &\hspace{0.5in} 
	&  $2*3*N^3$ \\
Zero modes $a^3_-(x_\bot):$  &\hspace{0.5in} &  $N^2$ \\
Residual Gauss' law constraints:  &\hspace{0.5in} &  $- N^2$  \\
 &\hspace{0.5in} & ------------------\\
&\hspace{0.5in} &  $6N^3$ \\
\hline
{\bf Final ${\cal H}$} & &  \\
$A_1^{ \prime a}, A_2^{ \prime a}(x_\bot,x_-):$  &\hspace{0.5in} 
	&  $2*3*N^3-(N^2-1)$ \\
Zero modes $a^3_-(x_\bot):$  &\hspace{0.5in} &  $N^2$ \\
Global neutrality constraint:   &\hspace{0.5in}
	&  $-1\,$ 	\\
\hspace{0.5in} &&------------------ \\
&\hspace{0.5in} &  $6N^3$	\\
\hline
\end{tabular}
\end{center}
\end{table}

The formulation of axial gauge near light front QCD for $SU(2)$
gauge fields is complete at this point.
For details of the rather subtle derivation we refer to
original references \cite{LNT,LNT2}. See also \cite{Thies} for
a pedagogical review and \cite{G} for a critical discussion focusing
on topological aspects.  In \cite{Jurek} the method using unitary
transformations has been
applied to QED quantized on the exact light cone.
The above  Hamiltonian already is rather complex and nonlocal, just as the
most familiar example of a gauge fixed theory, Coulomb gauge 
QED.  Despite the complexity it serves as a promising starting 
point for further studies since approximations can be made without
breaking local gauge invariance.

\section{Zero Mode Dynamics}

The principal advantage of an exact light front formulation is the apparent
triviality of the ground state which simplifies 
calculations of the hadron spectrum. 
The light front vacuum, however,  is not guaranteed to be trivial in the zero
mode sector. In using the near light front coordinate system,
we can study the complex zero mode structure
influencing  the dynamics  of
long distances. The zero mode sector in light front physics 
differs from equal time Hamiltonian physics where the
long range physics is low energy physics. As can be seen
from the dispersion relation for massless particles on the light front,
$p_+ = \frac{p^2_{\bot}}{2p_-},$
soft modes (states with small momenta) become high energy states.
In this way, high energy physics
becomes tied to long range physics, contrary to the equal time
formulation.
This physics appears in deep inelastic 
scattering at small scaling variable and is related to the long 
distance features of the proton.
We will focus on the zero  mode sector in order to
try to acquire some insight into its dynamics.  Therewith
we hope to obtain a sound  basis for further numerical studies.

From the comparison of abelian and non-abelian
theories, striking differences show up in the zero mode 
sector.
Recently, in the equal time formalism, the zero mode sector in
QCD has been
claimed to be relevant for the confinement phenomenon \cite{LMT}.
On the level of approximations and restrictions followed below,
the formal differences between light front and equal time 
approach are rather small and, consequently,  results and methods are similar.

In this  work 
we do not restrict ourselves to the strong coupling
approximation. We will, however, start with the strongly coupled
theory to define our set of basis functions.
As before, we will restrict ourselves to $SU(2)$ without matter --
pure gluonic Yang-Mills theory. It already has the typical non-abelian
features such as the Coulomb term which explicitly contains the zero
modes in the denominator and the non-standard kinetic energy for
the zero modes. 

The zero mode degrees of freedom couple to the three--dimensional
gluon fields  via
the second  term in ${\cal H}$ shifting the longitudinal
momenta of the transverse
gluon fields (Eq. (\ref{Hamil})). They affect  the Coulomb term
and  the two--dimensional electric fields ${e}_{\bot}$.
The latter coupling is typical for the light front and is absent in the
equal time case. We neglect these couplings and 
consider the pure zero mode Hamiltonian 
\begin{equation}
h = \int d^2x \; \left[ 
 \frac{1}{2 L}  p^{3\,\dagger}_-({x}_{\bot}) p^{3}_-({x}_{\bot})  
+ \frac{L}{2\eta^2} (\nabla_{\bot} a^3_-({x}_{\bot}))^2\right].
\label{Hamil3}
\end{equation}
This Hamiltonian is obtained from Eq. (\ref{Hamil}) with only
the zero mode $a^3_-$ and its conjugate momentum retained, and an
integration over the longitudinal variable.
We recognize `electric' and `magnetic' contributions in $h$, the zero mode 
Hamiltonian -- the first and second term, respectively. 
The light front variables mix the ordinary spatial and time  
variables so
the labeling above is to be understood in analogy with the equal time 
Hamiltonian. 

Even at this level of approximation, this zero mode
Hamiltonian differs from the corresponding one in QED. The reason 
is the
hermiticity defect of the canonical momentum: 
$p_-^{\dagger} \neq  p_-$.  This might seem to be a strange property 
for a momentum
operator, but is perfectly allowable, in analogy
with the  naive radial Schr\"odinger momentum operator which is also
non--Hermitian.

For notational simplicity, we now omit the color index and work
with the Schr\"odinger representation of Eq. (\ref{Hamil3})
\begin{equation}
h =  \int d^2x_{\bot} \; \left[ 
 -\frac{1}{2L} \frac{1}{J(a_-({x}_{\bot}))}
\frac{\delta}{\delta a_-({x}_{\bot})}   
 J(a_-({x}_{\bot})) \frac{\delta}{\delta a_-({x}_{\bot})}   
+ \frac{L}{2 \eta^2} (\nabla_{\bot} a_-({x}_{\bot}))^2\right] \; ,
\label{Hamil4}
\end{equation} 
where $J(a_-)$ is the Jacobian
and equals the Haar measure of 
$SU(2)$
\begin{equation}
 J(a_-({x}_{\bot})) = \sin^2 (\frac{gL}{2}a_-({x}_{\bot})).   
\end{equation}
The Jacobian is connected to the hermiticity defect of $p_-$;
they stem from the gauge fixing procedure taking into account the
curvilinear coordinates. The measure also appears in the
integration volume element for calculating matrix elements.
For ease of calculation, 
we introduce dimensionless variables 
\begin{equation}
\varphi({x}_{\bot}) = \frac{gL}{2}a_-({x}_{\bot}) \; ,
\end{equation}
in which Hamiltonian and Jacobian respectively read
\begin{equation}
h =  \int d^2x_{\bot} \; \left[ 
 -\frac{g^2 L}{8} \frac{1}{J(\varphi({x}_{\bot}))}
\frac{\delta}{\delta \varphi({x}_{\bot})}   
 J(\varphi({x}_{\bot})) \frac{\delta}{\delta 
\varphi({x}_{\bot})}   
+ \frac{2}{\eta^2 g^2 L} (\nabla_{\bot} \varphi({x}_{\bot}))^2\right] \; ,
\label{Hamil5}
\end{equation}
\begin{equation}
 J(\varphi({x}_{\bot})) = \sin^2 (\varphi({x}_{\bot})).   
\end{equation}
As in earlier approaches, see e.g. \cite{Bronz}, $\varphi$ will be
treated as a compact variable, $ 0 \le \varphi < \pi$. 

At this stage it is necessary to appeal to the physics of the infinite
momentum frame to factorize the reduced true energy $h_{\rm red}$ 
and the
Lorentz boost factor $\frac{\gamma}{\sqrt{2}} = \frac{1}{2 \eta}$, since
essentially $h$ is a light front energy, and it is well known how
these behave under a Lorentz transformation.  
We rewrite 
\begin{equation}
h = \frac{1}{2\eta} h_{\rm red} \, ,
\end{equation}
with
\begin{equation}
h_{\rm red} = \int d^2 x_\bot\left[-\frac{g^2 L\eta }
{4}\frac{1}{J}\frac{\delta}{\delta\varphi} J\frac{\delta}
{\delta\varphi}+\frac{4}{g^2 L\eta}(\nabla_\bot\varphi)^2\right].
\end{equation}
It should be noted that
the coefficients of the two terms are reciprocals of each other.

Since the integral over
transverse coordinates can contain arbitrarily small wavelengths, 
we have to regularize the above Hamiltonian $h_{\rm red}$.
We do this by introducing a lattice to evaluate the transverse 
integral. The lattice vector $\vec b$ numbers the lattice sites, 
and $\vec \varepsilon_{1}$ and $\vec \varepsilon_{2}$  
are the two unit vectors on the two--dimensional lattice.
In order to have standard commutation relations on the lattice 
the derivative on the lattice becomes
$\frac{\delta}{\delta\varphi_{\vec b}}=\frac{\delta}{\delta\varphi
(x_{\bot}) }a^2$.
We further explicitly pull out the
dependence on the lattice cutoff by defining a new reduced Hamiltonian
$\hat h_{\rm red}$ and substituting $\eta=\frac{1}{\sqrt 2} a M$, where
$M$ is a typical hadronic mass (see, e.g. \cite{tjthesis}):  
\begin{equation}
h=\frac{1}{2\eta a}\hat h_{\rm red} \, ,
\end{equation}
with
\begin{equation}
\hat h_{\rm red} = \sum_{\vec{b}}\left\{-\frac{g^2 LM}{4\sqrt2}
\frac{1}{J}
\frac{\delta}{\delta\varphi_{\vec b}} J\frac{\delta}{\delta\varphi 
_{\vec b}}
+\left(\frac{4\sqrt2}{g^2LM}\right) \sum_{\vec{\varepsilon}}
(\varphi_{\vec b}-\varphi_{\vec b+\vec\varepsilon})^2\right\} \, .
\end{equation}
Since the
effective coupling constant, 
\begin{equation}
g^{2}_{\rm eff}= \frac{g^2LM}{4\sqrt{2}} \, ,
\end{equation}
contains the large factor $LM$, the product of lattice size in the
longitudinal direction and the hadron mass, a strong coupling 
approach seems to be a good starting point. 

We separate the Hamiltonian into
electric and magnetic contributions:
\begin{equation}
\hat h_{\rm red}=\sum_{\vec b}\hat h^{\rm e}_{\vec b}+\sum_{\vec b}
\hat h^{\rm m}_{\vec b},
\label{hred}
\end{equation}
where
\begin{equation}
\hat h^{\rm e}_{\vec b}=-g^{2}_{\rm eff}
\frac{1}{J}
\frac{\delta}{\delta\varphi_{\vec b}} J\frac{\delta}{\delta
\varphi_{\vec b}},\end{equation}
and
\begin{equation}
\hat h^{\rm m}_{\vec b}=\frac{1}{g^{2}_{\rm eff}} \sum_{\vec{\varepsilon}}
(\varphi_{\vec b}-\varphi_{\vec b+\vec\varepsilon})^2.\end{equation}
Note that we do not introduce `radial wave functions'
nor effective potentials as in \cite{LNT,LMT}.
For each lattice site $\vec b$, the 
electric part of the Hamiltonian $\hat h^{\rm e}_{\vec b}$
(the kinetic energy) has the Gegenbauer
polynomials $C_{n_{\vec{b}}}(\varphi_{\vec b})$ as eigenfunctions:
\begin{equation}
\hat h^{\rm e}_{\vec b} C_{n_{\vec{b}}}(\varphi_{\vec b})=
g^{2}_{\rm eff}
n_{\vec b}(n_{\vec b}+2)
C_{n_{\vec{b}}}(\varphi_{\vec b}) , \end{equation}
with
\begin{equation}
C_{n_{\vec{b}}}(\varphi_{\vec b})=\sqrt{\frac{2}{\pi}}\left\{\frac{
\sin((n_{\vec b}+1)\varphi_{\vec b})}
{\sin\varphi_{\vec b}}\right\} , 
\label{Basis}
\end{equation}
and
\begin{equation}
\int^\pi_0  J(\varphi) C_{n}(\varphi) C_{m}(\varphi_)d\varphi
=\delta_{n m} \; .
\end{equation}
The strong coupling  wave functions of the full 
transverse lattice are product states
characterized by a set of quantum numbers $\{n\} = 
\{n_{\vec{b}}\}$,
\begin{equation}\label{23}
\Psi_{\{n\}}(\varphi)=\prod_{\vec b} C_{n_{\vec{b}}}(\varphi_{\vec b}) \, .
\end{equation}
These functions form a complete and orthonormal basis for the zero
mode sector.
They satisfy the energy eigenvalue equation
\begin{equation}
\sum_{\vec b}\hat h^{\rm e}_{\vec b}\Psi_{\{n\}}(\varphi)=
g^{2}_{\rm eff}\sum_{\vec b} n_{\vec b}(n_{\vec 
b}+2)\Psi_{\{n\}}(\varphi).\end{equation}
The ground state in this limit corresponds to all
$n_{\vec{b}} = 0$ -- a constant wave function 
\begin{equation}
\Psi_{\{0\}}(\varphi)=
\prod_{\vec{b} } \sqrt{\frac{2}{\pi}} \; ,
\end{equation}
and the ground
state energy is zero  
\begin{equation}
E_{0}=0.
\end{equation}
The first excited energy level is $N_{\bot}^2$-fold degenerate - an excitation
at a single lattice point
\begin{equation}
\Psi_{\{1\}}(\varphi)= \sqrt{\frac{2}{\pi}}
\frac{\sin\left( 2\varphi_{\vec{b}}\right)} {\sin\varphi_{\vec{b}}}
\prod_{\vec{b}' \ne \vec{b}} \sqrt{\frac{2}{\pi}} \; .
\end{equation}
In strong coupling this level is separated by a large amount from the ground 
state energy
\begin{equation}
E_{1}=3 g^2_{\rm eff} .
\end{equation}
So far our results are equivalent to those of \cite{LMT} to within
re--definitions of wave functions and integration measures.
In  \cite{Bronz} weak coupling variational   
solutions for the
full $SU(2)$ lattice Hamiltonian are given. Furthermore, 
studies in (1+1)--dimensional Yang-Mills theory \cite{Shaba} 
give formal extensions to construct wave functions  for
$SU(N)$ gauge theories.

The magnetic term of the Hamiltonian  couples nearest neighbor 
lattice points. In the strong coupling limit its contribution may be 
obtained perturbatively (as it has the coefficient $1/g_{\rm eff}^2$)
by evaluating it with the basis function of the ground state.
The result of this is  
\begin{equation}
\langle\Psi_{\{0\}} |\sum_{\vec b} h^{\rm m}_{\vec b}|\Psi_{\{0\}}\rangle
=\frac{1}{g^2_{\rm eff}}
\left(\frac{\pi^2}{6}-1\right)\cdot(2N_\bot^2) \, .\end{equation} 
Since this energy is proportional to  $ N^2_{\bot} = (L/a)^2$ and
$g_{\rm eff}^2$ grows linearly with $L$,
this part of the zero mode dynamics represents a negligible surface effect
for the three--dimensional system in the strong coupling approximation. 

Next,  we discuss  
the weak coupling limit $g^2\to 0$. In this case 
we can simplify the kinetic 
term of the
Hamiltonian by defining new variables $\alpha_{\vec b}$:
\begin{equation}
\alpha_{\vec b}=\frac{\varphi_{\vec b}} {\kappa g} \, ,
\end{equation}
with
$ 8 \kappa^2= \sqrt2 L M \, .$
Then the reduced Hamiltonian becomes:
\begin{equation}
\hat h_{\rm red}=\sum_{\vec
b}\frac{-1}{\sin^2(\kappa g \alpha_{\vec b}
))}\frac{\partial}{\partial\alpha_{\vec b} }\sin^2(\kappa g \alpha_{\vec b}
)\frac{\partial}{\partial
\alpha_{\vec b}}
+\sum_{\vec 
b,\vec\varepsilon}
(\alpha_{\vec b}(\vec b)-\alpha_{\vec b+\vec\varepsilon})^2.
\end{equation}
Expanding this for small $g$, we obtain
\begin{eqnarray}
\hat h_{\rm red}&=&
\sum_{\vec b} \left \{
-\left(\frac{\partial^2}{\partial \alpha_{\vec b}^2}
+\frac{2}{\alpha_{\vec b} }\frac{\partial}{\partial \alpha_{\vec b}
}\right)
+\sum_{\vec\varepsilon}(\alpha_{\vec b}-\alpha_{\vec b + \vec \epsilon}
)^2 \right \} .
\label{spin}
\end{eqnarray}

The eigensolutions of this Hamiltonian are known to be spin waves. 
Going over to 
Fourier momentum representation,
\begin{equation}
\alpha_{\vec b}=\sum_{\vec k} e^{i\vec k\vec b}R_{\vec k} \, ,
\end{equation}
with $k_i= 2\pi n_{i} / N_{\bot} a $ and $ n_{i}=0,\pm 1,\pm 2,\ldots$
we have
\begin{equation}
\hat h_{\rm red}=\sum_{\vec k}\left \{
- \left(\frac{\partial^2}{\partial
R^2_{\vec k}}+\frac{2}{R_{\vec k}}\frac{\partial}{\partial R_{\vec k}}\right)+4
\sum_{\vec\varepsilon}\sin^2\frac{\vec k\vec\varepsilon}{2}
R_{\vec k} R_{-\vec k}\right\} \, .
\end{equation}
The eigensolutions $\psi_K$ of $\hat h_{\rm red}$
in the weak coupling approximation are decoupled harmonic 
oscillators
for each $\vec k$, with frequencies
\begin{equation}
\omega^2_{\vec k }=4\sum_{\vec\varepsilon}\sin^2\frac{\vec 
k \vec\varepsilon}{2}.
\end{equation}
Because of the `radial Laplacian' it looks as if the eigenfunctions
would have to vanish at the origin to be normalizable. However,
as in the Schr\"odinger equation in three dimensions,
the Jacobian $J$  allows a constant wave function at the origin.
Consequently, the eigenvalue of $\psi_K$
is given by the sum over the modes:
\begin{equation}
\Omega_K=\sum_{\vec 
k}\sqrt{4\sum_{\vec\varepsilon}\sin^2
\left(\frac{\vec k\vec\varepsilon}{2}\right)} \, ,
\end{equation}
which gives in the $N_{\bot} \to\infty$ limit spin waves with  
$\omega_{\vec k}
=\sqrt{k^2_1+k^2_2}$.

In the weak coupling limit the zero mode Hamiltonian supports 
solutions similar to QED. The strong coupling limit, however,
yields different results: `gluonic'
excitations are suppressed because of large energy gaps.
This is due to the Jacobian, which can be traced back
to non-abelian self interactions in the original Lagrangian.

\section{Two--site truncation}

We have obtained solutions in both the weak and strong 
coupling regime.  We will now study the problem in the 
intermediate region.  As an initial effort, we will not solve the problem
for the full lattice.  Rather, we will calculate with what is  
essentially a cluster expansion \cite {tjthesis}, and we will start with the simplest, 
two--site cluster, in which either site (or both) can be excited to high 
energy states.  We will obtain the solution for the low--lying spectra
of the system approximated as a low density of excitable two--site
clusters.  This method can be envisaged as the starting point of a more
ambitious Hamiltonian based renormalization group technique, like the
contractor renormalization group method (CORE)\cite {MW}, or \cite{tjheff}.

We handle the calculation of the energies via an effective
Hamiltonian method. 
We work in the representation of the strong coupling solution  
of $\hat{h}^{\rm e}_{\vec b} $ and
divide the  two--site subspace into a $P$ and $Q$ space, such that 
$P+Q=1$ with 
\begin{eqnarray}
P&=&\{|0,0\rangle\}\, ,\nonumber\\
Q&=&\{|n,m\rangle;\ n, m \not= 0,0\}\,  ,
\end{eqnarray}
where $n,m$ represent the indices of the Gegenbauer polynomials.
Note that
we have picked our $P$ space as the strong coupling two--site 
ground state.
Then the two--site energy $E_2$ is given by the
non-perturbative solution of the Hamiltonian in Eq. (\ref{hred}),
truncated to two lattice sites.  Explicitly, this Hamiltonian is:
\begin{equation}
\hat h_{2}=\hat h^{\rm e} + \hat h^{\rm m},
\end{equation}
with
\begin{equation}
\hat h^{\rm e} = -g^{2}_{\rm eff}\left \{\frac{1}{J}\frac{\delta}{\delta\varphi_1} 
J\frac{\delta}{\delta\varphi_1} +
\frac{1}{J}\frac{\delta}{\delta\varphi_2} 
J\frac{\delta}{\delta\varphi_2}
\right \} ,
\end{equation}
and
\begin{equation}
\hat h^{\rm m}=\frac{1}{g^{2}_{\rm eff}} (\varphi_{1}-\varphi_{2})^2,
\end{equation}
where the subscripts label the sites. 

Within the effective Hamiltonian 
method, the two--site energy is given by (see, for example \cite{zvb}):
\begin{equation}
E_2=P\hat h_{2} P+ P\hat h_{2} Q\frac{1}{E_2-Q\hat 
h_{2} Q}
Q\hat h_{2} P \, .
\label{blocheq}
\end{equation}
The self--consistent solutions of this
equation provide the low--lying spectra in this method.
The strong coupling basis states are eigenstates of $\hat h^{\rm e}$:
\begin{equation}
\hat h^{\rm e} |n,m \rangle = g^2_{\rm eff}\left \{ n(n+2)+m(m+2)
\right \} |n,m \rangle  \, .
\label{unpertener}
\end{equation}
Thus, the non--trivial matrix elements are those of $\hat h^{\rm m}$,
and are of the form
\begin{equation}
\langle n,m | (\varphi_1-\varphi_2)^2 |n',m' \rangle \, .
\end{equation}

Although our states are two--site states, the operators appearing in the
matrix elements are simple one--site operators, and thus we can consider
the states to be products of one--site states.  
This reduces the evaluation to sums and
products of one--site matrix elements, which are given as:

\begin{eqnarray}
\langle n|\varphi| n'\rangle &=& \frac{\pi}{2} \hspace{4.3cm}{\rm for}\
n=n'  \nonumber \\
\langle n|\varphi| n'\rangle &=&\left\{\begin{array}{cl}
\frac{2}{\pi}\left(\frac{1}{(n+n'+2)^2}-\frac{1}{(n-n')^2}\right)
&{\rm for}\  n+n'= {\rm odd}\\
0&{\rm for}\  n+n'= {\rm even}, n \neq n'
\end{array}\right.  \nonumber\\
\langle n|\varphi^2| n' \rangle&=&\left\{\begin{array}{ll}
\frac{2}{\pi}\left[\frac{\pi^3}{6}-\frac{\pi}{[2(n+1)]^2}\right]
&{\rm for} \ n=n'\\
\frac{2}{\pi}\left\{\pi(-1)^{n+n'}\left[\frac{1}{(n-n')^2}
-\frac{1}{(n+n'+2)^2}\right]\right\}
&{\rm for}\  n\not= n'.\end{array}\right.
\end{eqnarray}
Before performing any numerical calculation with the effective 
Hamiltonian, it will be
illuminating to investigate the strong and weak coupling limits of
this theory. 

In the strong coupling limit we expect to obtain 
the result of perturbation theory in $1/g_{\rm eff}$.
The energy in this limit is easily calculated:
\begin{equation}
E_{2}=\langle 0,0|\hat h^{\rm m}|0,0\rangle= \frac{1}{g^2_{\rm eff}} \left \{
\langle 0 | \varphi^2 |0  \rangle  - 2
\left( \langle 0 | \varphi |0  \rangle \right )^2 \right \}
= \frac{1}{g^2_{\rm eff}} 
\left(\frac{\pi^2}{6}-1\right).
\label{2pert}
\end{equation}

In the weak coupling limit we can solve the Schr\"odinger equation for two 
neighboring sites - the Hamiltonian will simply be the two--site version
of the earlier spin wave Hamiltonian, Eq. (\ref{spin}).
We call the respective variables $\alpha_{\vec b_1}=x$ 
and $\alpha_{\vec b_2}=y$ ,
then we have to find the eigenenergies of the Hamiltonian:
\begin{eqnarray}
\hat h_{\rm red}&=&
-\left(\frac{\partial^2}{\partial x^2}
+\frac{2}{x}\frac{\partial}{\partial x}\right)
-\left(\frac{\partial^2}{\partial y^2}
+\frac{2}{y}\frac{\partial}{\partial y}\right) + (x-y)^2.
\end{eqnarray}
It should be noted that this Hamiltonian is invariant under $x
\leftrightarrow y$, and thus the eigenfunctions $\Psi_2(x,y)$
can be chosen to be symmetric under the interchange of $x$ and $y$ 
($\Psi_{2s}(x,y)$), or antisymmetric under 
the interchange of $x$ and $y$ ($\Psi_{2a}(x,y)$). 
Each of these symmetric or antisymmetric sets of solutions form a 
`tower' of excitations. 
The first symmetric excited state
becomes degenerate with the ground state of the original problem in the 
weak coupling limit, and the first antisymmetric state has a greater 
energy than the first symmetric state. 

As usual one factorizes the wave function
\begin{equation}
\Psi_2(x,y)=\frac{1}{xy}\Phi_2(x,y) \, ,
\end{equation}
resulting in the Schr\"odinger equation
\begin{equation}
\hat h_{\rm red} \Phi_2(x,y) = \left \{ -\frac{\partial^2}{\partial x^2}
-\frac{\partial^2}{\partial y^2} + (x-y)^2 \right \} \Phi_2(x,y) \,.
\end{equation}
The center--of--mass motion is then separated:
\begin{equation}
\Phi_2(x,y)=e^{iPR}\chi_2(r) \, ,
\end{equation}
with $R=(x+y)/2$ and $r=x-y$.
The Hamiltonian corresponding to the relative motion ($r$) is a simple
radial harmonic oscillator:
\begin{equation}
(\hat h_{\rm red})_{r}=-2\frac{\partial^2}{\partial r^2} + r^2.
\label{radial}
\end{equation}
The lowest states of the symmetric and antisymmetric `towers' are solutions 
to this Hamiltonian.
The energies of these states can be read directly from Eq.
(\ref{radial});
$ E_{2s}=\sqrt{2}$, and $ E_{2a}=3 \sqrt{2},$
respectively, 
giving a energy gap between the states of $2 \sqrt{2}$.

Thus, the results for the energy gaps of the low--lying states in the
weak coupling limit are
\begin{eqnarray}
E_{2s}-E_{\rm ground} &=& 0 \, ,\nonumber \\
E_{2a}-E_{\rm ground} &=& 2 \sqrt{2}  \, .
\label{analgap}
\end{eqnarray}

We now proceed to calculate the low--lying spectra 
via the effective Hamiltonian method. In the numerical calculations, we
cannot keep all states in the $Q$ space -- our choice is to cut at a
high two--site energy, calculate $E_2$, then increase the size of the
$Q$ space to check for convergent results.  This procedure was carried
out for each choice of
coupling constant $g_{\rm eff}$, 
and the typical number of two--site states kept in the $Q$ space
at convergence was about 300. 

\begin{figure}[hbt]
\epsfig{figure=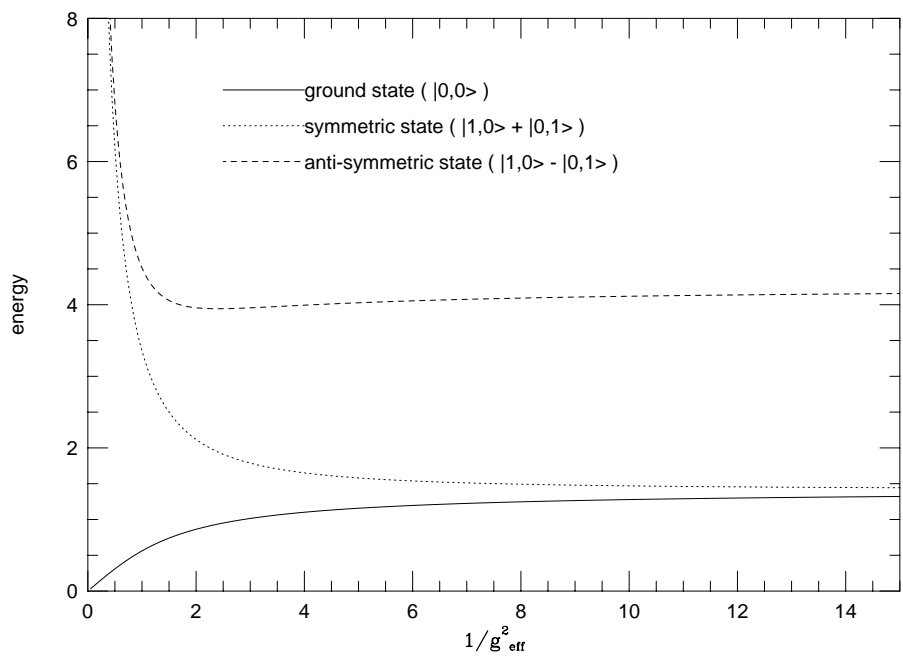, scale=0.5}
\caption{\label{states} Energies of first 3 states in the zero mode sector
calculated via effective Hamiltonian method.}
\end{figure}

The numerical solution of 
Eq. (\ref{blocheq}) for $E_2$ is given in Figure \ref{states}.
In the strong coupling limit (large  $g^2_{\rm eff}$) the large gaps in 
energy are evident, and the numbers agree with the unperturbed energy of
the states, given in Eq. (\ref{unpertener}).
In this same limit, the slope of the ground state energy 
as a function of
the inverse square coupling agrees with the analytic calculation 
of Eq. (\ref{2pert}).
In the weak coupling limit, the results for the gap energies were 
Richardson extrapolated 
for the $1/g^2_{\rm eff} \to \infty$ limit.  This extrapolation 
matched the analytical results of Eq. (\ref{analgap}) to five 
significant figures.  Thus, we have obtained two--site solutions
for the entire range of coupling which agreed with analytic results in
the weak and strong coupling limits.  

It is remarkable that we succeeded a one--dimensional strong
coupling basis throughout the range of coupling strengths. 
Results for the spectra of $N$--sites   
are straightforwardly obtained as long as the number of excited
two--site clusters is small compared with $N/2$.  This is the `low
density' approximation.

\section{Effect of Zero Modes on Confinement}

In order to study the confinement problem, we start with the
assumption that the three--dimensional gauge Hamiltonian
can be treated in a weak coupling approximation  with 
small gauge coupling $g^2$, whereas the physics of the 
two--dimensional zero mode subsystem can be obtained
in a strong coupling approximation for $g^2_{\rm eff}=
\frac{g^2L M}{4 \sqrt2}$.
We will see that the asymmetric treatment
of transverse and longitudinal spatial
coordinates has effects, which go 
beyond the violation of rotational symmetry in strong coupling
lattice gauge theory. They are connected with the procedure of
initially choosing an axial gauge in three dimensions and then letting
a two--dimensional Coulomb gauge follow. 
One advantage of this approach is that 
the dynamical role played 
by the zero modes becomes particularly illuminating. 
The zero modes preserve
the axial linear confinement in first order perturbation theory.
Thus, we argue that already at the level of
the zero mode sector of $SU(2)$ pure glue QCD we see evidence
for confinement.  

Let us demonstrate the usefulness of the basis functions obtained
in the strong coupling approximation, cf. Eq. (\ref{Basis}),
by considering  matrix elements of the Coulomb 
term.
In terms of the variables $\varphi$ defined on a discretized
transverse space the Coulomb potential explicitly reads
\begin{equation}
H_C  =  
 \frac{La^2}{4} \sum_{\vec{b}} \int_{0}^{L} dz^{-} \int_{0}^{L} dy^{-}
\sum_{p,q,n}\,'\frac{ G'_{\bot qp}(\vec{b}, z^{-})
G'_{\bot pq}(\vec{b},y^{-})}{\left[ \pi n +
(p-q)\varphi_{\vec{b}} \right]^{2}}
e^{i2\pi n(z^{-}-y^{-})/L} \ .
\label{Hcou1}
\end{equation}
In order to separate possible singularities we work out the $p, q$ 
sum:
\begin{eqnarray}
H_C & = & 
 \frac{La^2}{4} \sum_{\vec{b}} \int_{0}^{L} dz^{-} \int_{0}^{L} dy^{-}
\sum_{n \ne 0}\,\frac{ G'_{\bot 11}(\vec{b}, z^{-}) G'_{\bot 
11}(\vec{b},y^{-})
+ G'_{\bot 22}(\vec{b}, z^{-}) G'_{\bot 22}(\vec{b},y^{-})}
{\left[ \pi n  \right]^{2}}
e^{i2\pi n(z^{-}-y^{-})/L} \nonumber \\
& + & 
 \frac{La^2}{4} \sum_{\vec{b}} \int_{0}^{L} dz^{-} \int_{0}^{L} dy^{-}
\sum_{n}\,\frac{ G'_{\bot 12}(\vec{b}, z^{-}) G'_{\bot 21}(\vec{b},y^{-
})}
{\left[ \pi n + \varphi_{\vec{b}} \right]^{2}}
e^{i2\pi n(z^{-}-y^{-})/L} \nonumber \\
& + & 
 \frac{La^2}{4} \sum_{\vec{b}} \int_{0}^{L} dz^{-} \int_{0}^{L} dy^{-}
\sum_{n}\,\frac{ G'_{\bot 21}(\vec{b}, z^{-}) G'_{\bot 12}(\vec{b},y^{-
})}
{\left[ \pi n - \varphi_{\vec{b}} \right]^{2}}
e^{i2\pi n(z^{-}-y^{-})/L} \, .
\label{Hcou2}
\end{eqnarray}
Obviously the `abelian' terms with  $G'_{11}$ and $G'_{22}$ 
are infrared regular since $n \ne 0$ in that case. The non-abelian 
terms can have singularities for $\varphi \rightarrow 0, \pi$.  
Exploiting the strong coupling basis, we show that 
these terms are also infrared regular. Such a  `dynamical 
regularization' is anticipated, because of the connection of
the non-standard zero mode kinetic energy, 
the hermiticity defect, and the Coulomb term: the Jacobian vanishes
at the points, where the `propagator' becomes singular.
Consider the last two terms in Eq. (\ref{Hcou2}),
which can be  added using 
$z\leftrightarrow y$, $n\leftrightarrow -n$. This yields the non-abelian
Coulomb contribution $H_c$:
\begin{equation}
H_c=\frac{La^2}{2}\sum_{\vec b}\int^L_0
dz^-\int^L_0 dy^-\sum_n\frac{G'_{\bot 21}(\vec b,z^-)G'_{\bot 12}
(\vec b,y^-)}{(\pi n-\varphi_{\vec b})^2} e^{2\pi in(z^--y^-)/L}.
\end{equation}
In this expression, we can identify the `Coulomb propagator' 
$D_{c}(z^-- y^-,\varphi_{\vec b})$ in position
space, and it can be evaluated with the result \cite {ES}
\begin{eqnarray}
D_{c}(z^-- y^-,\varphi_{\vec b})& \equiv &\frac{L}{2}\sum_n
\frac{e^{2\pi in(z^--y^-)/L}}{(\pi n-\varphi_{\vec 
b})^2}\nonumber\\
&=&e^{2i\varphi_b\cdot(z^--y^-)/L}\left[\frac{L}{2\sin^2\varphi_{\vec b}}
-|z^--y^-|
-i(z^--y^-)\cot\varphi_{\vec b}\right]\nonumber\\
& \equiv & D_1+D_2+D_3.\end{eqnarray}
In the continuum limit one finds additional
terms besides the linear propagator in one dimension. In the strong 
coupling
approximation we integrate this Coulomb propagator and the 
off-diagonal $G'_{\bot 21}G'_{\bot 12}$ with the Gegenbauer 
polynomials $C_0(\varphi)$ over
$d\varphi J(\varphi)$ appropriate for the curvilinear coordinates.
At this point it is evident that the Jacobian indeed
prevents possible infrared singularities mentioned above.

Let us discuss  the first term $D_1$ 
in the Coulomb propagator.
With the strong coupling ground state 
$\Psi_{\{0\}}$
it leads to a Coulomb energy 
\begin{eqnarray}
\langle \Psi_{\{0\}}| H_c^{(1)} |\Psi_{\{0\}} \rangle
&=&\langle\Psi_{\{0\}}|a^2 \sum_{\vec b}\int^L_0 dz^-\int^L_0
dy^-\ G'_{\bot 21}(\vec b,z^-)G'_{\bot 12}(\vec b,y^-) D_1 |\Psi_{\{0\}} 
\rangle\nonumber\\
&=&L a^2\sum_{\vec b}\int^L_0 dz^-\int^L_0 dy^-
G'_{\bot 21}(\vec b,z^-)G'_{\bot 12}(\vec b,y^-) \times \nonumber\\
&&\quad\left\{\frac{L}{2\pi i(z^--y^-)}\left[ e^{2\pi i(z^--y^-)/L}
-1\right]\right\}\nonumber\\
&&\longrightarrow L a^2\sum_{\vec b}|\tilde{Q}_{12}(\vec b)|^2,\end{eqnarray}
for large $L$, with 
\begin{equation}
\tilde{Q}_{12}(\vec b)=\int^L_0 dz^- G'_{\bot 12}({\vec b},z^-).
\end{equation}
One sees that the expectation value of the
off-diagonal `charge' $\tilde{Q}_{12}(\vec b)$ at each
transverse site $\vec b$ has to vanish in order to avoid an infinite
Coulomb energy in the continuum limit
\begin{equation}
\langle \tilde{Q}_{12}(\vec b) \rangle \equiv0\qquad \forall \vec b.
\end{equation}
Recall that, in order to fully resolve Gauss' Law,
the third component of the global charge should vanish
in the physical sector (cf. Eq. (\ref{Gaussgl})).
These two conditions together 
suggest that the physical states are 
real color singlets not merely color
neutral states with a color three projection equal zero.
The second term $D_2$ of the Coulomb propagator 
gives a linearly rising potential in the 
$L\rightarrow\infty$ limit, 
\begin{eqnarray}
\langle \Psi_{\{0\}}| H_c^{(2)} |\Psi_{\{0\}} \rangle&=&
\langle \Psi_{\{0\}}|a^2\sum_{\vec{b}}\int^L_0 dz^-\int^L_0 dy^-\ G'_{\perp 21}(\vec 
b,z^-)
G'_{\perp 12}(\vec b,y^-)D_2| \Psi_{\{0\}} \rangle \nonumber\\
& \rightarrow &  - a^2\sum_{\vec{b}}\int^L_0 dz^-\int^L_0 dy^-\
G'_{\perp 21}(\vec b,z^-)G'_{\perp 12}(\vec b,y^-)|z^--y^-| \, .
\end{eqnarray}
Neglecting terms proportional to $g$
with gluons in $G'_{\perp 21}$ and only considering the 
external charges
one gets a confining linear potential. Two color spin 1/2
point charges coupled to a color singlet at the same 
transverse site interact in longitudinal direction with the potential:
\begin{equation}
V_{12} = 
\frac{-1}{a^2}\frac{g^2}{4}\langle\vec\tau_1
\vec\tau_2\rangle |z^--y^-|=\frac{3}{4}g^2|z^--y^-|\frac{1}{a^2} \, .
\end{equation}
The scale of the string tension is given in strong coupling by the lattice size
$a$ of the transverse lattice. In the continuum calculation it should be 
replaced  by a correlation length generated in  
the transverse zero mode dynamics.
The third term 
in the Coulomb propagator $D_3$ does not contribute to the strong
coupling ground state energy in the $L \rightarrow \infty$ limit, 
since then $ \langle \tilde{Q}_{12}(\vec b) \rangle =0$.

One of the 
main advantages of the light front Coulomb gauge or axial gauges for QCD 
is visible here.
Whereas in QED the Coulomb gauge is designed to give
the $1/r$ potential, the above gauge choices give confining 
potentials in zeroth order. 
Naturally, these potentials are linked to the choice of
gauge, so one has to consider the perturbative corrections to the
gauge potential. In axial gauge QED the potential for opposite charges has 
the form
\begin{equation}
M^{QED}_1=g^2|z^--y^-|\frac{1}{a^2} \, ,
\end{equation}
which appears to have the same confining properties as the QCD result.
However, the first-order one photon exchange correction 
for soft photons  with  momenta $(q_-,\vec q_\perp)$ has a spin
independent 
contribution 
\begin{equation}
M^{QED}_2=g^2\frac{\vec
q^2_\perp}{q_-q_-}\frac{\theta(q_-)}{q_-}\frac{-1}{\vec q^2_\perp/q_-}.
\end{equation}
After transforming $M^{QED}_2$ to coordinate space, it is seen that this 
one-photon exchange cancels the
confining gauge artifact and  the Coulomb potential 
$1/r$ plus spin
dependent corrections remain. 

In QCD the one-gluon exchange contributions 
have to be discussed separately for color charged gluons 
$A_\perp^{\prime 1,2}(x^-,x_\perp)$ and color neutral gluons.
The $A_\perp^{\prime 1,2}$ components of the transverse 
gluon fields interact with the zero mode $a_-^3(x_\perp)$ fields.
These `background fields' have  non--zero expectation values
in the strong coupling limit for $g_{\rm eff}$.
In this way 
the $A^{\prime 1,2}_\perp$ fields may acquire a mass. In other words, we
argue that the dispersion relation of 
transverse colored gluons possibly is changed to
\begin{equation}
k_+=\frac{1}{\eta^2}
\left(-k_- +\sqrt{k_{-}^{2}+\eta^2(k_{\perp}^2+\frac{4}{\eta^2}
\langle \varphi^2 \rangle /L^2)}\right).
\end{equation}
Consequently, the one--gluon exchange can no longer cancel the linear
confinement potential at large distances, as the fields are now massive
and thus finite--range. 

Our approach produces similar results as the 
similarity transformation scheme \cite{WW,P}. 
This scheme has been 
proposed by Glazek and Wilson \cite {GW} for the light front Hamiltonian 
and by Wegner \cite {W}
independently for condensed matter physics. It avoids  vanishing energy 
denominators for the 
$q_-\to0$ region for one--gluon exchange by a cutoff 
$\lambda$. In  higher orders the region of validity for $q_-$ can be 
enlarged successively (the cutoff $\lambda$ can be made smaller). In 
our approach
near the light front we can give the physical origin of the cutoff 
$\lambda$. It
lies in the zero mode fields $a_-(x_\perp)$ which limit 
the long--range
propagation of gluon fields. We only use the strong coupling 
approximation for
the ground state wave functional of the transverse zero mode 
lattice. Certainly 
higher orders and a more accurate description of the zero mode 
dynamics are necessary
to prove that the linear confinement potential is preserved.
Color neutral gluon fields $A^{3^\prime}_\perp(x^-,x_\perp)$ do not 
directly interact with $a^3_-(x_\perp)$, but by their construction in
Eq. (\ref{subt}) their two--dimensional
longitudinal zero modes ($q_-=0$) have  been subtracted. The corresponding
two-dimensional transverse parts, however,
are still present as dynamical modes.

It is more complicated to analyze  
two color charges separated in transverse direction.
Consider two color charges separated by $x_\perp$  and
oriented along the color 3-direction. 
As color 3-charges they experience the
two-dimensional Coulomb potential via the electric field
\begin{equation}
e_\perp^{3}(x_\perp) = g \nabla_{\perp }\int dy^- 
dy_\perp \, d(x_\perp -
y_\perp) \rho^3_m(y^-, y_\perp).
\end{equation}
The above propagator generates a logarithmic 
potential, which is
also confining. This potential is unaffected by 
the zero modes $a_-$, since the  strong coupling
ground state expectation
value of $\nabla_{\bot}a_-$ vanishes.
Explicitly, we obtain
\begin{equation}
V_{12}=\frac{g^2}{a^2 M^2 L}\frac{\tau_1^3 \tau_2^3}{4}ln 
|\frac{x_{\bot}-y_{\bot}}{a}|.
\end{equation}
It corresponds to spreading flux lines in 2 dimensions and strongly 
violates the rotational invariance
in the spatial coordinates. 
To investigate this problem further, one must understand 
the role played by the zero modes in the continuum limit.
Work in this direction is in progress \cite {HJP}.

\section{Summary}

We have presented a near light front description of QCD.
In our Hamiltonian approach, formulated in a finite volume,
the modified light-cone gauge $\partial_-A_-=0$ is a natural choice.
The resulting  asymmetric treatment of transverse and longitudinal dynamics 
matches the physics of deep inelastic scattering where
large momenta of the hadron are involved.
We argue that the well--known physical appeal of an 
infinite momentum frame  
can indeed be realized rather naturally in near light 
front coordinates.  

In the chosen canonical formulation, Gauss' law
appears as a quantum mechanical constraint to be
implemented on physical states. It reflects the presence
of redundant gauge variables. Using unitary gauge fixing
transformations, these are eliminated by implementing
Gauss' law. The result is a Hamiltonian formulated
in terms of unconstrained dynamical variables.

This Hamiltonian contains zero mode degrees of freedom, which
only depend on transverse coordinates.
The zero-mode dynamics apparently 
generates the dominant non--perturbative physics with
interesting implications. 
In the strong coupling limit,
$g^2_{\rm eff}=\frac{g^2LM}{4\sqrt2} \gg 1$, 
the energy gaps between excited `gluonic' states and the
ground state are large -- in contrast to QED.
Another indication of confinement is a linear
quark-quark potential in longitudinal direction, which
is not merely a gauge artifact -- as it is in QED.
In transverse direction
confinement is supported via a logarithmic quark-quark potential.
One has yet to discover how the two-dimensional dynamics relates to
the full-three dimensional dynamics.

The eigenfunctions of the zero mode Hamiltonian in the weak
coupling limit are spin waves. We also have investigated the
intermediate coupling regime by means of a cluster expansion.
Via an effective Hamiltonian method the two-site energy
was explicitly calculated for the entire range of couplings.
We demonstrated that the results indeed converge to the
analytical expressions for
$g^2_{\rm eff} \gg 1$  and $g^2_{\rm eff} \ll 1$ .
The spectra of $N$-sites can be constructed in the `low density'
approximation.

Our approach opens a potential path 
to construct constituent quark models on the light front 
by eliminating the negative energy solutions from the near
light front Hamiltonian.
The effective Hamiltonian method utilized here for the zero modes may
also be the appropriate tool to study
the mechanism of chiral symmetry breaking. 

\setcounter{equation}{0}
\renewcommand{\theequation}{A.\arabic{equation}}
\section*{Appendix: $SU(N)$ Hamiltonian including fermions }
\appendix
The resolution of the Gauss law constraint via `unitary gauge
fixing transformations' can also be achieved in the $SU(N)$-case,
including dynamical fermions \cite{LNT,LNT2}.
The final Hamiltonian in  near light front coordinates is formulated 
in terms of the following
$SU(N)$ variables:
\begin{itemize}
\item{Fermion fields $\psi$ and $\psi^{\dagger}$, obeying standard anti-commutation
relations. (The $N$ color indices, as well as Dirac and flavor labels
are suppressed.) It should be emphasized that
$\psi_- = \frac{1}{2}(1-\alpha^3)\psi$ is a dynamical
field, whereas on the exact light--front it is constrained.}
\item{Transverse gluon fields, $A_i^{'a} (i=1,2 ; a=1,....,N^2-1)$,
without the neutral, two-dimensional, longitudinal parts, which have been
eliminated. This means that
\begin{equation}
\partial_i \int_{0}^{L} dx^{-} A_i^{' c_0}(\vec{x}_{\bot}, x^-) = 0,
\end{equation}
where the color index with subindex '0' refers to the Cartan subalgebra.
In the usual representation this corresponds to diagonal $\lambda$-matrices.
The conjugate chromoelectric fields are denoted with $\Pi_i^{'a}$.}
\item{Zero mode gluon fields $a_-$ and their canonical momenta $p_-$. 
Neither of these fields depend on $x^-$ and both are color diagonal. 
Thus one can write
\begin{eqnarray}
a_-(\vec{x}_{\bot}) &=& \sum_{c_0=1}^{N-1} a^{c_0}_-(\vec{x}_{\bot})
\frac{\lambda^{c_0}}{2}\, ,
\nonumber \\
p_-(\vec{x}_{\bot}) &=&  \sum_{c_0=1}^{N-1}p^{c_0}_-(\vec{x}_{\bot})
\frac{\lambda^{c_0}}{2}\, .
\end{eqnarray}
}
\end{itemize}
The Hamiltonian density in the physical Hilbert space explicitly reads
\begin{eqnarray}
{\cal H} & = & - i \frac{2}{\eta^2} \psi^{\dagger}_-
 \left( \partial_{-}-iga_{-}\right) \psi_-
-i \frac{1}{\eta}\psi^{\dagger}\vec{\alpha}_{\bot}
 \left( \vec{\nabla}_{\bot} - ig \vec{A}'_{\bot} \right)  \psi +
m  \frac{1}{\eta}\psi^{\dagger} \beta \psi \nonumber \\
& + &  \mbox{tr} \left[  \partial_1 A'_2 -\partial_2 A'_1
-ig [A'_1, A'_2]  \right]^2 
 +  \frac{1}{\eta^2} \mbox{tr} \left[ 
\vec{\Pi'_{\bot}}-\left(\partial_-\vec{A}'_{\bot}-
ig[a_-,\vec{A}'_{\bot}]\right)\right]^2 \nonumber \\
& + &  \frac{1}{\eta^2} \mbox{tr} \left[ \frac{1}{L} \vec{e}_{\bot} -
\nabla_{\bot} a_-\right]^2 
 +  \frac{1}{2 L^{2}} \sum_{c_{0}} p^{c_{0} \dagger}_{-}(\vec{x}_{\bot})
p^{c_{0}}_{-}(\vec{x}_{\bot})  \nonumber \\
& + &   \frac{1}{L^{2}}\int_{0}^{L} dz^{-} \int_{0}^{L} dy^{-}
\sum_{p,q,n}\,'\frac{ G'_{\bot qp}(\vec{x}_{\bot}, z^{-})
G'_{\bot pq}(\vec{x}_{\bot},y^{-})}{\left[ \frac{2\pi n}{L} +
g(a_{-,q}(\vec{x}_{\bot})- a_{-,p}(\vec{x}_{\bot})) \right]^{2}}
e^{i2\pi n(z^{-}-y^{-})/L} ,
\end{eqnarray}
with $p, q = 1, ....,N$.
The operator $G'_{\perp}$,  the  neutral chromo-electric fields
  $\vec{e}_{\bot},$ and the neutral charge $Q^{c_0}$, which
appears below, are defined as
the generalizations of the corresponding quantities in the main text.
Note that the matter density contains the dynamical fermion fields, 
\begin{equation}
\rho_m^{a}(\vec{x}_{\bot}, x^-) = \psi^{\dagger}(\vec{x}_{\bot}, x^-)
\frac{\lambda^a}{2} \psi(\vec{x}_{\bot}, x^-) \, .
\end{equation}
The variables defined above are not constrained
apart from the global conditions
\begin{equation}
Q^{c_0} | \Phi' \rangle = 0 .
\end{equation}
The neutral components of the  total color-charge vanish
in the sector of (transformed) physical states. 
This completes the $SU(N)$ formulation of axial gauge light cone QCD.

\end{document}